\documentclass [12pt]{article}
\usepackage {a4wide}
\usepackage {graphicx}
\usepackage {longtable}

\begin{document}
\begin{center}
\textbf{Energy and Momentum in Expansive Nondecelerative Universe}
\end{center}

\bigskip

\begin{center}
Miroslav S\'{u}ken\'{\i}k and Jozef \v{S}ima 
\end{center}

\begin{center}
Faculty of Chemical Technology, Slovak Technical University, Radlinsk\'{e}ho 9, 
812 37 Bratislava, Slovakia
\end{center}

\begin{center}
\underline {sima@chtf.stuba.sk}
\end{center}

\bigskip

\textbf{Abstract.} Incorporation of the Vaidya metric in the model of 
Expansive Nondecelerative Universe allows to precisely localize 
gravitational energy for weak fields and obtain the components of the 
Einstein energy-momentum pseudotensor for strong gravitational fields. The 
components are identical to those calculated by Virbhadra.

\bigskip

\subsection*{Introduction}

\bigskip

It is becoming still more obvious that the solution of several problems of 
cosmology and astrophysics depends on the ability to localize gravitational 
energy. In this area three main streams of opinions can be identified: 1) 
gravitational energy is localizable but a corresponding "magic" formula for 
its density is to be found; 2) gravitational energy is nonlocalizable in 
principle; 3) gravitational energy does not exist at all since the 
gravitational field is a pure geometric phenomenon. 

One of the approaches to the localization of gravitational energy is based 
on a presumption that space-time may have more than four dimensions. 
Kaluza-Klein theories [1] seem to be the first to have proposed a solution 
within this approach. The corresponding D-dimensional metric has the form

\begin{equation}
\label{eq1}
ds^{2} = g_{\mu \nu}  \left( {x^{\mu} } \right)dx^{\mu} dx^{\nu}  - \gamma 
_{ab} \left( {x^{a}} \right)dx^{a}dx^{b}
\end{equation}

\noindent
where $g_{\mu \nu}  $ is the metric of four-dimensional world while $\gamma 
_{ab} $ is the metric associated with $\mbox{D} - 4$ compact extra dimensions.

During 80's the theory of superstrings was developed, a number of dimensions 
in it reaching 10. An interesting solution comes from six-dimensional space 
model, where the gravity is localized on a four-dimensional singular 
string-like defect [2]. A six-dimensional metric satisfying 4D Poincar\'{e} 
invariance is as follows

\begin{equation}
\label{eq2}
ds^{2} = \sigma \left( {\rho}  \right)g_{\mu r} dx^{\mu} dx^{\nu}  - d\rho 
^{2} - \gamma \left( {\rho}  \right)d\theta ^{2}
\end{equation}

For two extraspatial dimensions the polar coordinates $\rho ,\theta $ were 
introduced. A different approach dealing with gravitational energy-momentum 
density in teleparallel gravity has led to a true space-time and gauge 
tensor transforming covariantly under global Lorentz transformation [3].

Attempts to localize the gravitational energy in a classic four-dimensional 
space-time face the problem of scalar curvature that, using Schwarzschild 
metric, is of zero value outside a body. The way out may lie in application 
of a different metric consistent with a new model of the Universe, the model 
of Expansive Nondecelerative Universe (ENU) [4]. 

The present paper is devoted to the solution of Einstein pseudotensor and, 
in turn, to localization and quantification of the density of gravitational 
field energy in four-dimensional space-time.

\bigskip

\begin{center}
\subsection*{Background of the ENU model}
\end{center}

\bigskip

The model of Expansive Nondecelerative Universe [4] differs from more 
frequently used inflation models in the following features:

\noindent
a) Schwarzschild metric is replaced by Vaidya metric, 

\noindent
b) the Universe permanently expands by the velocity of light $c$,

\noindent
c) simultaneous creation of matter and the equivalent amount of 
gravitational energy (which is, however, negative and thus the total value 
of mass-energy is constant and equal zero) occurs. Statements b) and c) can 
be expressed as follows

\begin{equation}
\label{eq3}
a = c.t_{c} = \frac{{2G.m_{U}} }{{c^{2}}}
\end{equation}

\begin{equation}
\label{eq4}
\Lambda = 0
\end{equation}

\begin{equation}
\label{eq5}
k = 0
\end{equation}

\noindent
where $a$ is the gauge factor, $m_{U} $ is the Universe mass, $\Lambda $ is 
the cosmological member, $k$ is the curvature index, and $t_{c} $ is the 
cosmological time. Introducing dimensionless conform time, equation (\ref{eq3}) can 
be expressed as
\begin{equation}
\label{eq:eq6a}
c.dt = a.d\eta
\end{equation}

\noindent
from which

\begin{equation}
\label{eq6}
a = \frac{{da}}{{d\eta} }
\end{equation}

Friedmann equations can be then written in the form [5]

\begin{equation}
\label{eq7}
\frac{{d}}{{d\eta} }\left( {\frac{{1}}{{a}}.\frac{{da}}{{d\eta} }} \right) = 
- \frac{{4\pi .G}}{{3c^{4}}}a^{2}\left( {\varepsilon + 3p} \right)
\end{equation}

\begin{equation}
\label{eq8}
\left( {\frac{{1}}{{a}}.\frac{{da}}{{d\eta} }} \right)^{2} = \frac{{8\pi 
.G}}{{3c^{4}}}a^{2}\varepsilon - k
\end{equation}

\noindent
where $\varepsilon $ is the energy density and $p$ is the pressure. Based on 
(\ref{eq7}) and (\ref{eq8}) it follows

\begin{equation}
\label{eq9}
\varepsilon = \frac{{3c^{4}}}{{8\pi .G.a^{2}}}
\end{equation}

\begin{equation}
\label{eq10}
p = - \frac{{\varepsilon} }{{3}}
\end{equation}

Equations (\ref{eq9}) and (\ref{eq10}) represent the matter creation and the negative value 
of gravitational energy, respectively. It can be evidenced that in the ENU 
it holds [6]

\begin{equation}
\label{eq11}
\sum \frac{{dm}}{{dt}} = \frac{{dm_{U}} }{{dt}} = \frac{{m_{U}} }{{t_{c} 
}} = \frac{{c^{3}}}{{2G}}
\end{equation}

\noindent
and

\begin{equation}
\label{eq12}
m_{U} = \frac{{a.c^{2}}}{{2G}}
\end{equation}

One of the corner-stones of the ENU lies in a nonstatic nature of 
space-time, understood here as a time change in the Universe mass, $dm_{U} 
/dt$ (\ref{eq11}). Solutions of some astrophysical problems have taken this nature 
into account (e.g. its importance for treatment of the problem of naked 
singularities [7] or radiating stars [8, 9] was clearly manifested. A 
contribution of the ENU is in the fact that instead of a general expression 
of $dm_{U} /dt$ it offers its value as

\begin{equation}
\label{eq13}
\frac{{dm}}{{dt}} = \frac{{m}}{{t_{c}} }
\end{equation}

Based on (\ref{eq13}) it is obvious that metric with varying Newton potential, in 
which the scalar curvature is of nonzero value also outside of body, should 
be applied. The first to solve this problem was Vaidya dealing with radiant 
stars. 

The density of gravitational energy $\varepsilon _{g} $ is in the first 
approximation expressed by Tolman equation [10]

\begin{equation}
\label{eq14}
\varepsilon _{g} = - \frac{{R.c^{4}}}{{8\pi .G}}
\end{equation}

\noindent
where $R$ is the scalar curvature. Using Vaidya metric and applying relation 
(\ref{eq13}) we obtain [11,12]

\begin{equation}
\label{eq15}
R = \frac{{6G}}{{c^{3}.r^{2}}}.\frac{{dm}}{{dt}} = \frac{{3r_{g} 
}}{{a.r^{2}}}
\end{equation}

\noindent
where $r_{g} $is the gravitational radius of a body with the mass $m.$ 
Relations (\ref{eq14}) and (\ref{eq15}) lead to the following expression for the density of 
gravitational field energy

\begin{equation}
\label{eq16}
\varepsilon _{g} = - \frac{{3m.c^{2}}}{{4\pi .a.r^{2}}}
\end{equation}

The total amount of gravitational energy emitted by a body in time unit 
(gravitational output $P_{g} $) is given by

\begin{equation}
\label{eq17}
P_{g} = - \frac{{d}}{{dt}}\int {\frac{{c^{4}.R}}{{8\pi .G}}dV = - 
\frac{{m.c^{2}}}{{t_{c}} }} 
\end{equation}

In the ENU, the wavefunction of the Universe is formulated [11] as

\begin{equation}
\label{eq18}
\Psi _{g} = e^{ - i\left( {t_{Pc} .t_{c}}  \right)^{ - 1/2}.t}
\end{equation}

\noindent
where $t_{Pc} $ is the Planck time. For the energy of gravitational 
quantumit follows [12] from (\ref{eq18})

\begin{equation}
\label{eq19}
\left| {E_{g}}  \right| = k.T = \left( {\frac{{\hbar ^{3}.c^{5}}}{{t_{c}^{2} 
.G}}} \right)^{1/4}
\end{equation}

\noindent
where $T$ is the Universe temperature. Relation (\ref{eq19}) documents that the 
density of gravitational energy and that of radiation energy were identical 
during the entire radiation era, i.e. the Universe had to be in 
thermodynamic equilibrium up to the end of the radiation era.

\bigskip

\subsection*{Energy-Momentum in ENU}

\bigskip

In the Vaidya metric [8,9] the line element is formulated in the form

\begin{equation}
\label{eq20}
ds^{2} = \frac{{\Psi '^{2}}}{{f_{\left( {m} \right)}^{2}} }\left( {1 - 
\frac{{2\Psi} }{{r}}} \right).c^{2}.dt^{2} - \left( {1 - \frac{{2\Psi 
}}{{r}}} \right)^{ - 1}dr^{2} - r^{2}\left( {d\theta ^{2} + sin^{2}\theta 
.d\varphi ^{2}} \right)
\end{equation}

\noindent
where

\begin{equation}
\label{eq21}
\Psi = \frac{{G.m}}{{c^{2}}}
\end{equation}

It follows from (\ref{eq13}) and (\ref{eq21}) in the ENU

\begin{equation}
\label{eq22}
\Psi ' = \frac{{d\Psi} }{{c.dt}} = \frac{{\Psi} }{{a}}
\end{equation}

 $f_{\left( {m} \right)} $ is an arbitrary function. In order it has a nonzero 
value, it must hold in the ENU

\begin{equation}
\label{eq23}
f_{\left( {m} \right)} = \Psi \left[ {\frac{{d}}{{dr}}\left( {1 - 
\frac{{2\Psi} }{{r}}} \right)} \right] = \frac{{2\Psi ^{2}}}{{r^{2}}}
\end{equation}

Gravitational influence can be realized in the ENU only when the absolute 
value of gravitational energy density will exceed the critical energy 
density. Thus, if (\ref{eq9}) and (\ref{eq16}) are equal, then

\begin{equation}
\label{eq24}
R_{ef} = \left( {R_{g} .a} \right)^{1/2} = \left( {2\Psi .a} \right)^{1/2}
\end{equation}

\noindent
where $R_{ef} $ is the effective gravitational range of a body with the 
gravitational radius $R_{g} .$ Vaidya metric is applicable in the ENU if

\begin{equation}
\label{eq25}
r < R_{ef} 
\end{equation}

It follows from (\ref{eq20}) that scope of Vaidya metric applicability is determined 
by relation

\begin{equation}
\label{eq26}
\Psi ' = f_{\left( {m} \right)} 
\end{equation}

Based on (\ref{eq22}), (\ref{eq23}) and (\ref{eq26}) it holds

\begin{equation}
\label{eq27}
A = \frac{{c^{4}}}{{4\pi .G}}
\end{equation}

Equation (\ref{eq27}) represents the condition of changing Vaidya metric to 
Schwarzschild metric in which the scalar curvature becomes of zero value 
which prevents to localize gravitational energy. This conclusion conforms 
with the ENU model.

The energy-momentum complex of Einstein pseudotensor adopts the form [7]

\begin{equation}
\label{eq28}
\theta _{i}^{k} = \frac{{1}}{{16\pi} }\left[ {\frac{{g_{in}} }{{\sqrt { - g} 
}}\left\{ { - g\left( {g^{kn}.g^{lm} - g^{ln}.g^{km}} \right)} \right\},_{m} 
} \right]_{,_{l}}  
\end{equation}

Vaidya metric (\ref{eq20}) in Cartesian Kerr-Schild coordinates was solved by 
Virbhadra [8] who found the following components of Einstein pseudotensor 
(in the Virbhadra's notation, geometrized units in which the speed of light 
and Newtonial gravitational constant are taken $c = G = 1$). For the sake of 
simplicity, we use a simplified notation where:

\begin{equation}
\label{eq29}
A = \frac{{c^{4}}}{{4\pi .G}}
\end{equation}

\begin{equation}
\label{eq30}
\alpha = \frac{{1}}{{r^{3}}}
\end{equation}

\begin{equation}
\label{eq31}
\beta = \frac{{\alpha} }{{r}}
\end{equation}

\bigskip

Then the pseudotensor components are as follows:

\begin{equation}
\label{eq32}
\begin{array}{l}
 \theta _{0}^{0} = - \frac{{A.\Psi '}}{{r^{2}}} \\ 
 \\ 
 \end{array}
\end{equation}

\begin{equation}
\label{eq33}
\theta _{1}^{0} = - \theta _{0}^{1} = Ax\alpha \Psi '
\end{equation}

\begin{equation}
\label{eq34}
\theta _{2}^{0} = - \theta _{0}^{2} = Ay\alpha \Psi '
\end{equation}

\begin{equation}
\label{eq35}
\theta _{3}^{0} = - \theta _{0}^{3} = Az\alpha \Psi '
\end{equation}

\begin{equation}
\label{eq36}
\theta _{1}^{1} = A\beta x^{2}\Psi '
\end{equation}

\begin{equation}
\label{eq37}
\theta _{2}^{2} = A\beta y^{2}\Psi '
\end{equation}

\begin{equation}
\label{eq38}
\theta _{3}^{3} = A\beta z^{2}\Psi '
\end{equation}

\begin{equation}
\label{eq39}
\theta _{1}^{2} = \theta _{2}^{1} = A\beta xy\Psi '
\end{equation}

\begin{equation}
\label{eq40}
\theta _{2}^{3} = \theta _{3}^{2} = A\beta yz\Psi '
\end{equation}

\begin{equation}
\label{eq41}
\theta _{3}^{1} = \theta _{1}^{3} = A\beta xz\Psi '
\end{equation}

Our calculations stemming from the ENU provide the same components as the 
above found by Virbhadra. In case of weak gravitational fields, Tolman 
equation allows to localize the density of gravitational field energy 
precisely (\ref{eq16}). It is obvious that Einstein pseudotensor is compatible with 
Tolman equation and the ENU model. 

Virbhadra has manifested [12] general applicability of Vaidya metric. The 
present work is aimed to provide evidence on the applicability of Vaidya 
metric in solving the problems of cosmology. It seems to be obvious that 
further progress in the field of gravitational energy localization depends 
of the ability to formulate a true tensor including the changes in the 
Universe mass.

\bigskip

\subsection*{Acknowledgement}

\bigskip

The authors are indebted to Professor K.S. Virbhadra for reprints of his 
papers and valuable suggestions.

\bigskip

\subsection*{References}

\bigskip
\mbox{}

\indent
1. T. Appelquist, A. Chodos, P.G. Freund (Eds.), \textit{Modern Kaluza-Klein 
Theories}, Addison-Wesley, Reading MA, 1987

2. T. Gherghetta, M. Shaposhnikov, \textit{Phys. Rev. Lett., 85} (2000) 240

3. V.C. de Andrade, L.C.T. Guillen, J.G. Pereira, \textit{Phys. Chem. Lett., 
84} (2000) 4533

4. V. Skalsk\'y, M. S\'{u}ken\'{\i}k, \textit{Astrophys. Space Sci., 178} (1991) 169 

5. A.N. Monin,\textit{ Cosmology, Hydrodynamics and Turbulence: A. A. 
Friedmann and Extension of his Scientific Heritage,} Moscow (1989), p. 103 
(in Russian)

6. 6. V. Skalsk\'y, M. S\'{u}ken\'{\i}k, \textit{Astrophys. Space Sci., 181} (1991) 
153\textit{} 

7. K.S. Virbhadra, \textit{Phys. Rev., D60} (1999) 104041

8. K.S. Virbhadra, \textit{Pramana - J. Phys., 38} (1992) 31

9. P.C. Vaidya, \textit{Proc. Indian Acad. Sci., A33} (1951) 264

10. R.C. Tolman, \textit{Relativity, Thermodynamics and Cosmology,} Oxford 
University Press, Oxford, 1934

11. J. \v{S}ima, M. S\'{u}ken\'{\i}k, \textit{General Relativity and Quantum Cosmology,} 
Preprint gr-qc/9903090; gr-qc/0011060

12. M. S\'{u}ken\'{\i}k, J. \v{S}ima, J. Vanko, \textit{Acta Phys. Slov.} (in press)

13. I.L. Rozental, \textit{Adv. Math. Phys. Astronomy, 31} (1986) 241 (in 
Czech)

\end{document}